\documentclass[prb,aps,twocolumn,showpacs]{revtex4-1}

\usepackage{graphicx,color}
\usepackage{amsthm}
\usepackage{amsfonts}
\usepackage{algorithmic}
\usepackage{enumerate}
\usepackage{latexsym}
\usepackage{amsmath}
\usepackage{amssymb}
\usepackage{bm}
\usepackage[pdftex,plainpages=false,colorlinks=true,linkcolor=blue, citecolor=blue, urlcolor=blue]{hyperref}

\emergencystretch=\maxdimen
\begin{document}
\title{Anomalous Orbital Reconstruction Controlled by Interorbital Correlations and Hund's Coupling}

\author{Yu Ni$^{1,2}$}
\email{niyu@ynnu.edu.cn}
\author{Lvjin Wang$^{1,2}$}
\author{Jinfeng Wang$^{1,2}$}
\author{Yunkun Niu$^{3}$}
\email{ykniu@imu.edu.cn}
\author{Zhaoming Fu$^{1,2}$}
\email{fuzhm1979@163.com}
\author{Yun Song$^{4}$}
\email{yunsong@bnu.edu.cn}

\affiliation{$^{1}$College of Physics and Electronic Information, Yunnan Normal University, Kunming 650500, China}
\affiliation{$^{2}$Yunnan Key Laboratory of Opto-Electronic Information Technology, Kunming 650500, China}
\affiliation{$^{3}$School of Physical Science and Technology $\&$ Inner Mongolia Key
Laboratory of Microscale Physics and Atomic Manufacturing, Inner Mongolia University,
Hohhot 010021,  China}
\affiliation{$^{4}$School of Physics and Astronomy, Beijing Normal University, Beijing 100875, China}


\begin{abstract}

The microscopic origin of anomalous orbital polarization in a class of low-dimensional correlated oxides remains unresolved due to the competition among crystal-field effects, electronic correlations, and orbital-dependent dimensionality.
Using dynamical mean-field theory, we investigate a two-orbital Hubbard model
with orbital-dependent dimensionality and reveal the mechanism for anomalous orbital polarization through orbital reconstruction beyond the bare crystal-field
picture.
We identify interorbital Coulomb interaction-induced charge competition as a key microscopic mechanism responsible for the orbital redistribution, which leads first to an orbital-polarized correlated metal and subsequently to an orbital-polarized Mott insulator.
We further find that Hund's coupling acts as a filling-dependent regulator of orbital reconstruction. It weakens the correlation-induced orbital redistribution at quarter filling by competing with interorbital charge fluctuations, while at half filling it completely suppresses the orbital-polarized state by stabilizing high-spin orbital-balanced configurations.
These results provide a unified picture of correlation-driven orbital
reconstruction and highlight the relevance of interorbital interactions and
Hund's coupling for understanding orbital phenomena in low-dimensional
transition-metal oxides.

\end{abstract}


\maketitle


\section{Introduction}

The interplay among charge, spin, and lattice degrees of freedom in strongly correlated electron systems gives rise to a rich variety of quantum phases, including unconventional superconductivity, Mott metal-insulator transitions (MIT), and non-Fermi-liquid behavior\cite{Khomskii2020,Paschen2021,Li2021,Chen2025,Prichard2025}.
In multiorbital correlated systems, the additional orbital degrees of freedom introduce new complexity into the electronic structure, as the interplay among crystal-field splitting, orbital-dependent hopping, and local electronic interactions provides multiple competing channels for determining the low-energy properties\cite{Georges2013Ann,Ko2023,Luo2023,Wang2024,Nguyen2025}.
Particularly important among these are the interorbital Coulomb ($U'$) interaction and Hund's coupling ($J_{\rm H}$), which play essential roles in controlling charge redistribution, electronic localization, and the competition between different correlated states\cite{Georges2013Ann,Ouyang2024,Aucar2024,Grundner2025,Bhardwaj2025}.
Understanding how electronic correlations modify orbital occupations and stabilize different correlated phases is therefore a central issue in condensed matter physics.

In multiorbital systems, orbital polarization, namely the preferential occupation of specific orbitals, plays a crucial role in determining the electronic properties of transition-metal oxides and oxide heterostructures\cite{Okamoto2004,Streltsov2017,Salluzzo2009,Yoshimatsu2010,Li2023,Fu2024}.
In conventional systems, orbital occupations are primarily determined by the single-particle crystal-field splitting, where electrons preferentially occupy orbitals with lower crystal-field energies\cite{Khomskii2020}.
However, recent experiments and theoretical studies have revealed anomalous orbital polarization in low-dimensional correlated oxides, where the observed orbital occupation cannot be explained by the bare crystal-field hierarchy alone\cite{Salluzzo2009,Yoshimatsu2010,Delugas2011,Beck2018,Green2021,Fu2024}.
In these systems, electrons can be redistributed toward orbitals that are energetically unfavorable in the noninteracting crystal-field picture, indicating that orbital-dependent kinetic energy and electronic correlations can reconstruct the conventional orbital hierarchy.
Such orbital reconstruction is particularly prominent in oxide interfaces and thin films, where coupled charge, spin, and orbital degrees of freedom generate emergent electronic states\cite{Hwang2012}.
For example, orbital reconstruction has been observed at LaAlO$_3$/SrTiO$_3$ interfaces, where interfacial confinement and electronic reconstruction modify the orbital occupation of the Ti $3d$ states\cite{Salluzzo2009}.
In addition, dimensional confinement in SrVO$_3$ thin films induces a strong orbital polarization toward the in-plane $d_{xy}$ orbital\cite{Yoshimatsu2010}.
These observations demonstrate that electronic correlations and reduced dimensionality can fundamentally modify the crystal-field picture. However, the microscopic mechanism by which correlations overcome the bare orbital hierarchy remains unclear. In particular, the respective roles of interorbital Coulomb interaction and Hund's coupling in driving orbital redistribution remain unresolved.

Recent theoretical work has attributed anomalous orbital polarization in low-dimensional oxides to the cooperative effects of dimensional confinement and electronic correlations\cite{Fu2024}.
Orbital-dependent hopping geometries produce distinct kinetic-energy scales, while correlations can reverse the orbital preference expected from the bare crystal-field splitting.
This picture explains why the higher crystal-field-energy in-plane orbital can become energetically favorable.
An unresolved issue, however, is why the orbital redistribution develops only beyond a finite interaction strength and can take the form of a sharp transition.
This suggests that, in addition to crystal-field renormalization, the evolution of interorbital many-body correlations must be considered.

Hund's coupling introduces a second competing interaction scale\cite{Held1998,Werner2007,Medici2009,Haule2009,Medici2011L,Georges2013Ann}.
Its effect is strongly filling dependent: it promotes high-spin Mott physics near half filling, whereas away from half filling it can suppress orbital fluctuations and substantially shift the localization boundary\cite{Medici2011L,Medici2011B}.
How this filling-dependent Hund's physics competes with anomalous orbital polarization in the presence of crystal-field splitting and orbital-dependent dimensionality has not yet been systematically clarified.

In this work, we investigate the competition among Coulomb interactions, crystal-field splitting, Hund's coupling, and orbital-dependent dimensionality in a two-orbital Hubbard model using dynamical mean-field theory (DMFT).
The {\it Lanczos} solver gives our computational scheme a clear advantage, enabling us to identify the interorbital charge-correlation channel responsible for the correlation-driven orbital reconstruction.
We demonstrate that anomalous orbital polarization originates from the interplay between orbital-dependent dimensionality and electronic correlations, with $U'$-driven interorbital charge competition providing the microscopic mechanism for orbital redistribution toward the higher energy in-plane orbital.
The suppression of interorbital double occupancy and the enhancement of negative interorbital charge correlation reveal the buildup of this competition before the orbital transition.
Furthermore, we show that orbital reconstruction precedes Mott localization, forming an orbital-polarized (OP) correlated metallic phase.
Hund's coupling acts as a filling-dependent regulator. It weakens the correlation-induced orbital redistribution at quarter filling by competing with interorbital charge fluctuations, while at half filling it completely suppresses the OP state by stabilizing high-spin orbital-balanced configurations.
These results establish a unified picture in which orbital-dependent dimensionality and crystal-field splitting define the bare orbital hierarchy, while electronic correlations and Hund's coupling determine the stability and evolution of reconstructed orbital phases.

The remainder of this paper is organized as follows.
In Sec.~II, we introduce the two-orbital Hubbard model and the DMFT method.
In Sec.~III, we present the numerical results for quarter-filled and half-filled systems and discuss the microscopic mechanism of orbital reconstruction and its dependence on Hund's coupling.
Finally, Sec.~IV summarizes our conclusions.

\section{model and method}

In quasi-two-dimensional oxides such as CaVO$_3$ thin films and LaAlO$_3$/SrTiO$_3$ interfaces, dimensional confinement differentiates the in-plane $d_{xy}$ orbital from the out-of-plane $d_{xz}/d_{yz}$ orbitals, leading to strongly orbital-dependent electronic structures\cite{Zhong2013,Fu2024}.
Although a full three-orbital description is necessary for a quantitative treatment of material-specific electronic structures, the essential orbital competition can be captured by an effective two-orbital model, where orbital 1 and orbital 2 represent the in-plane and out-of-plane channels, respectively.
Such reduced low-energy descriptions have been widely employed in correlated oxide interfaces and multiorbital Hubbard models\cite{Lechermann2014,Werner2007}.
The present minimal model retains the key ingredients responsible for orbital reconstruction, including orbital-dependent kinetic energy, crystal-field splitting, interorbital Coulomb interaction, and Hund's coupling.


The tight-binding (TB) part of the Hamiltonian, denoted  $H_t$, is given by
\begin{eqnarray}
    H_t&=&-\sum_{\left\langle ij \right\rangle}\sum_{l \sigma}
    (t^{x/y}_{l} d^{\dag}_{il\sigma}d_{jl\sigma}+c.c.)\nonumber\\
    &&+\sum_{il\sigma}(\epsilon_l-\mu) d^{\dag}_{il\sigma}d_{il\sigma}.
\end{eqnarray}
$d^{\dag}_{il\sigma}$ ($d_{il\sigma}$) is an electron creation (annihilation) operator for orbital $l$ at site $i$ with spin $\sigma$.
$t_l^x$ ($t_l^y$) denotes the nearest-neighbor hopping amplitude along the $x$ ($y$) direction, respectively.
Under Fourier transformation, the TB Hamiltonian is expressed as
\begin{eqnarray}
H_{t}(k)=\sum_{k l \sigma}
\xi_{l }(k) d_{l \sigma}^{\dagger}(k) d_{l^{\prime} \sigma}(k),
\label{Eq:H0}
\end{eqnarray}
with
\begin{eqnarray}
\xi_{l}(k) =-2 \left(t^{x}_{l}\cos k_{x}+t^{y}_{l}\cos k_{y}\right)+(\epsilon_l-\mu).
\label{Eq:DR}
\end{eqnarray}

In quasi-two-dimensional transition-metal oxides, dimensional confinement leads to distinct hopping dimensionalities: the in-plane $d_{xy}$ orbital can disperse along both directions, whereas the out-of-plane $d_{xz}/d_{yz}$ orbitals exhibit quasi-one-dimensional hopping.
To characterize this orbital-dependent dimensionality,
we assign the in-plane channel (orbital 1) hopping parameters $t^x_1=t^y_1=0.5$ eV, the out-of-plane channel (orbital 2) $t^x_2=0.5$ eV and $t^y_2=0$, and denote the on-site energy of orbital $l$ as $\epsilon_l$.
The crystal-field splitting is defined as $\epsilon_d=\epsilon_1-\epsilon_2=0.1$ eV.
Therefore, the out-of-plane orbital has a lower onsite energy than the in-plane orbital.

The hopping amplitude of the in-plane orbital is chosen as $t_1=0.5$ eV, which is comparable to the characteristic kinetic-energy
scale of $t_{2g}$ bands in transition-metal oxides such as SrVO$_3$, SrTiO$_3$ and LaAlO$_3$/SrTiO$_3$-based systems\cite{Yoshimatsu2010,Zhong2013,Tomczak2014,Fu2024}.
This choice provides a representative energy scale for comparison with correlated materials, while the present model is not intended as a quantitative description of a specific compound.
Instead, it captures the essential competition among orbital-dependent dimensionality, crystal-field splitting, orbital-dependent kinetic energy, interorbital Coulomb interaction, and Hund's coupling underlying correlation-driven orbital reconstruction.

The interaction Hamiltonian $H_I$ is exactly the same as the correlation part of the standard two-orbital Hubbard model\cite{Koga2005},
\begin{eqnarray}
    H_I&=&\frac{U}{2}\sum_{il\sigma}n_{il\sigma}n_{il\bar{\sigma}}+\sum_{i,l<l',\sigma\sigma'}
    (U'-\delta_{\sigma\sigma'}J_{\rm H})n_{il\sigma}n_{il'\sigma'}
    \nonumber\\
    &&+\frac{J_{\rm H}}{2}\sum_{i,l\neq l',\sigma} d^{\dag}_{il\sigma}d^{\dag}_{il\bar{\sigma}}d_{il'\bar{\sigma}}d_{il'\sigma}
    \nonumber\\
    &&+\frac{J_{\rm H}}{2}\sum_{i,l\neq l',\sigma\sigma'} d^{\dag}_{il\sigma}d^{\dag}_{il'\sigma'}d_{il\sigma'}d_{il'\sigma},
\label{Eq:TOHub}
\end{eqnarray}
where $U$ ($U'$) corresponds to the intraorbital (interorbital) interaction, and $J_{\rm H}$ is the Hund's coupling. For systems with spin rotation symmetry, we have $U=U'+2J_H$.
For transition-metal oxides, such as CaVO$_3$ and LaAlO$_3$/SrTiO$_3$ interfaces, the correlated $d$ electrons are typically described by local Coulomb interactions with $U\approx3\sim6$ eV and Hund's coupling $J_{\rm H}\approx0.5\sim0.7$ eV.
These interaction parameters have been widely adopted in previous studies of these low-dimensional transition-metal oxides\cite{Nekrasov2006,Lechermann2014,Beck2018}.
In the present work, we investigate a broader parameter range of $U=0\sim10$ eV and $J_{\rm H}/U=0\sim0.2$, which covers the realistic regime of Ti/V-based oxides and extends toward stronger correlated interaction and Hund's coupling.
This allows us to systematically elucidate the filling-dependent role of Hund's coupling in controlling correlation-driven orbital reconstruction and the resulting quantum phase transitions.

In the DMFT procedure, we map the lattice Hamiltonian onto an impurity
model with fewer degrees of freedom,
\begin{eqnarray}
{H_{imp}} &=& \sum\limits_{ml\sigma} {{\varepsilon_{ml\sigma}}} c_{ml\sigma}^\dag {c_{ml\sigma}}+ \mathop \sum \limits_{l\sigma} (\epsilon_l-\mu)d_{l\sigma}^\dag {d_{l\sigma}} \nonumber\\
&&+ \sum\limits_{lm\sigma} {{V_{lm\sigma}}\left( {d_{l\sigma}^\dag {c_{ml\sigma}} + c_{ml\sigma}^\dag {d_{l\sigma}}} \right)}  \nonumber\\
&& + {H_I^{imp}}\,
\label{Eq:IMP}
\end{eqnarray}
with
\begin{eqnarray}
    H_I^{imp}&=&\frac{U}{2}\sum_{l\sigma}n_{l\sigma}n_{l\bar{\sigma}}+\sum_{l<l',\sigma\sigma'}
    (U'-\delta_{\sigma\sigma'}J_{\rm H})n_{l\sigma}n_{l'\sigma'}
    \nonumber\\
    &&+\frac{J_{\rm H}}{2}\sum_{l\neq l',\sigma} d^{\dag}_{l\sigma}d^{\dag}_{l\bar{\sigma}}d_{l'\bar{\sigma}}d_{l'\sigma}
    \nonumber\\
    &&+\frac{J_{\rm H}}{2}\sum_{l\neq l',\sigma\sigma'} d^{\dag}_{l\sigma}d^{\dag}_{l'\sigma'}d_{l\sigma'}d_{l'\sigma},
\end{eqnarray}
where $c^{\dag}_{ml\sigma}$ ($c_{ml\sigma}$) denotes the creation (annihilation) operator for the $m$-th environmental bath lattice of orbital $l$,~$\varepsilon_{ml\sigma}$~denotes the energy of the $m$-th bath of orbital $l$, and $V_{lm\sigma}$ represents the coupling of the orbital $l$ between the impurity site and the $m$-th bath.
The Weiss function of the impurity model can be expressed in terms of the impurity Hamiltonian parameters,
\begin{eqnarray}
{\cal G}_0^{(l)}{\left( {i{\omega _n}} \right)^{ - 1}} = i{\omega _n} -(\epsilon_l-\mu)  - \sum\limits_{m} {\frac{{{V^2_{lm}}}}{{i\omega_n  - {\varepsilon_{ml}}}}},\
\end{eqnarray}

Employing the {\it Lanczos} solver, we obtain the diagonal matrix element of the Green's function $G^{imp}_{l}$\cite{Dagotto1994,Caffarel1994,Capone2007,YKNiu2019},
from which the self-energy matrix of the impurity Hamiltonian is determined by the Dyson equation,
\begin{eqnarray}
{\Sigma} _{imp}(i\omega _n) = {\cal G}_0^{(l)}{\left( {i{\omega _n}} \right)^{ - 1}}-G^{imp}_{l}\left( i \omega_n \right)^{ - 1}.\
\end{eqnarray}
Subsequently, the lattice Green's function takes the form
\begin{eqnarray}
{G}^{lat}_{l}\left( {i\omega _n} \right) &=& \frac{1}{N}\sum\limits_k {G^{(l)}_{ii}}\left( {i\omega _n,k} \right)  \nonumber\\
&=& \frac{1}{N}\sum\limits_k {\frac{1}{{i\omega _n - \xi_{l}(k) - {\Sigma} _{imp}(i\omega _n)}}}, \
\end{eqnarray}
where $\xi_{l}(k)$ is the dispersion relation of Eq.~(\ref{Eq:DR}).
Additional computational details are given in the Appendix~\ref{App-A}.

\section{calculated results}

\subsection{Correlation-induced orbital reconstruction in a mixed-dimensional Hubbard model}

\begin{figure}
\includegraphics[scale=0.43]{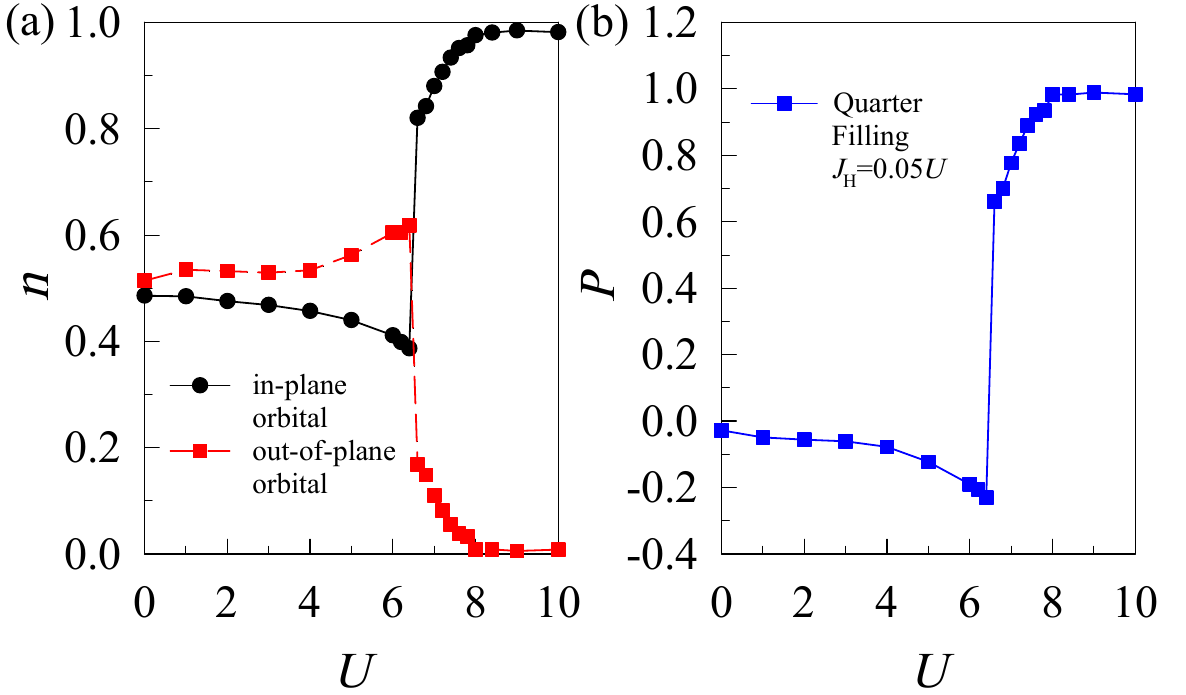}
\caption{Orbital reconstruction in the quarter-filled system with $J_{\rm H}=0.05U$.
(a) Orbital-resolved occupations as a function of the Coulomb interaction strength $U$.
In the weakly correlated regime, the occupation difference is mainly determined by the crystal-field splitting.
The occupation of the out-of-plane orbital remains dominant at weak coupling, followed by a sudden transfer of electrons into the in-plane orbital at $U\approx6.6$, leading to anomalous orbital polarization.
(b) Orbital polarization order parameter $P$ as a function of $U$.
The rapid increase and sign change of $P$ indicates the emergence of a correlation-driven orbital-reconstruction transition.
\label{n-U_P-U}}
\end{figure}

We first investigate the quarter-filled system, where the interplay between orbital-dependent dimensionality and electronic correlations drives an anomalous orbital reconstruction, manifested as a reversal of orbital polarization. By analyzing the evolution of orbital occupations and spectral properties, we identify how correlations overcome the bare crystal-field hierarchy and drive electrons toward the higher crystal-field-energy in-plane orbital.
To quantify the orbital imbalance resulting from this reconstruction, we define the orbital polarization order parameter as
\begin{equation}
P=\frac{n_1-n_2}{n_1+n_2},
\end{equation}
where $n_1$ and $n_2$ denote the occupations of the in-plane and out-of-plane orbitals, respectively.
Here, negative and positive values of $P$ correspond to the crystal-field-favored and anomalous orbital polarization states, respectively.

Figure~\ref{n-U_P-U} shows the evolution of the orbital occupations $n$ and polarization order $P$ with increasing Coulomb interaction strength $U$, and all energy parameters in this paper are expressed in units of eV.
At weak coupling, the orbital occupation is dominated by the bare crystal-field splitting, and the lower-energy out-of-plane orbital is preferentially occupied, resulting in a negative polarization  order parameter. This behavior is consistent with the conventional crystal-field picture.
However, this tendency is reversed when electronic correlations become sufficiently strong.
As shown in Fig.~\ref{n-U_P-U}(a), the occupation of the out-of-plane orbital gradually decreases, while electrons are transferred into the in-plane orbital.
Beyond $U\approx6.6$, the in-plane
orbital becomes preferentially occupied despite its higher crystal-field-energy, signaling the emergence of anomalous orbital polarization (the spectral-weight redistribution responsible for the nonmonotonic occupation behavior before the transition is discussed in Appendix~\ref{App-B}).
The reversal is clearly reflected in the orbital polarization order parameter shown in Fig.~\ref{n-U_P-U}(b). The sign change of $P$ demonstrates that the interaction-induced orbital reconstruction overcomes the bare crystal-field hierarchy and reverses the orbital preference.
Therefore, the anomalous orbital polarization observed here cannot be understood from the single-particle crystal-field splitting alone, highlighting the essential role of electronic correlations in reconstructing the orbital occupations.

\begin{figure}
\includegraphics[scale=0.40]{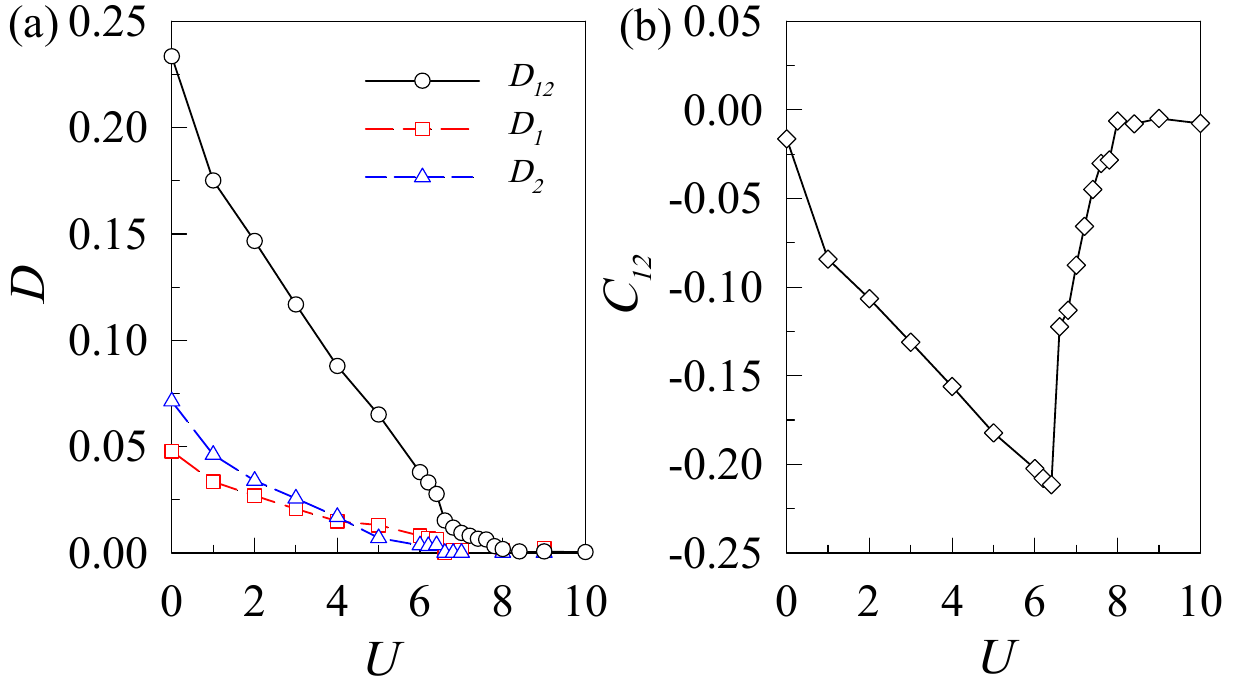}
\caption{Microscopic origin of correlation-driven orbital reconstruction at quarter filling with $J_{\rm H}=0.05U$.
(a) Evolution of intraorbital and interorbital double occupancies. $D_{12}$ denotes the probability of simultaneous occupation of the two orbitals, while $D_1$ and $D_2$ represent intraorbital double occupancies of the in-plane and out-of-plane orbitals, respectively.
The suppression of $D_{12}$ with increasing U indicates enhanced interorbital charge competition.
(b) Interorbital charge correlation function $C_{12}$.
The increasing negative correlation before the transition reveals strengthened interorbital charge competition associated with $U'$, which drives orbital redistribution.
\label{DC-U}}
\end{figure}

To identify the microscopic origin of the orbital-reconstruction transition, we calculate the local charge correlations.
To quantify the evolution of local charge fluctuations, we calculate the intraorbital and interorbital double occupancies
$D_1=\langle n_{1\uparrow}n_{1\downarrow}\rangle$,
$D_2=\langle n_{2\uparrow}n_{2\downarrow}\rangle$,
and
$D_{12}=\langle n_1n_2\rangle$.
Here $n_i=n_{i\uparrow}+n_{i\downarrow}$, $D_1$ and $D_2$ characterize the probability of double occupation within each orbital, while $D_{12}$ describes interorbital occupation.
We further introduce the interorbital charge correlation function
\begin{equation}
C_{12}=\langle n_1n_2\rangle-\langle n_1\rangle\langle n_2\rangle .
\end{equation}
Figure~\ref{DC-U}(a) shows that the interorbital double occupancy $D_{12}$ is progressively suppressed with increasing $U$, indicating a reduction of local configurations in which the two orbitals are simultaneously occupied.
More direct evidence for the competition between orbital occupations is provided by the connected interorbital charge correlation function $C_{12}$ in Fig.~\ref{DC-U}(b).
Before the orbital-reconstruction transition, $C_{12}$ becomes increasingly negative, indicating that the two orbitals develop stronger occupation anticorrelation. This enhanced interorbital charge competition reflects the increasing importance of the interorbital Coulomb interaction $U'$, which penalizes simultaneous occupation of different orbitals and favors orbital redistribution. Therefore, the simultaneous suppression of $D_{12}$ and enhancement of negative $C_{12}$ provide direct evidence for the buildup of a $U'$-driven interorbital charge competition preceding the orbital-reconstruction transition.
After the transition, $C_{12}$ approaches zero because the out-of-plane orbital becomes nearly occupied and the other nearly empty, thereby freezing the interorbital charge fluctuations rather than restoring an uncorrelated two-orbital state.

These results identify the microscopic origin of anomalous orbital polarization.
Rather than being determined solely by the bare crystal-field hierarchy, the orbital redistribution results from the interplay between orbital-dependent dimensionality and electronic correlations.
The dimensional confinement modifies the kinetic-energy balance and spectral structure of the two orbital channels, while the interorbital Coulomb interaction enhances charge competition between them.
This interaction-induced competition promotes electron redistribution toward the initially disfavored in-plane orbital, allowing the system to overcome the crystal-field preference for the lower-energy out-of-plane orbital.
Our results extend the dimensionality--correlation picture proposed in Ref.~\onlinecite{Fu2024} by further clarifying the microscopic role of interorbital correlations.
Dimensional confinement establishes the orbital-dependent kinetic-energy hierarchy, while correlation effects reconstruct the orbital occupations through enhanced interorbital charge competition at finite interaction strength, ultimately leading to the reversal of the crystal-field orbital hierarchy and polarization in higher crystal-field-energy in-plane orbital (the role of orbital-dependent dimensionality is further examined by comparing with an isotropic two-dimensional model in Appendix~\ref{App-C}).

\begin{figure}
\includegraphics[scale=0.60]{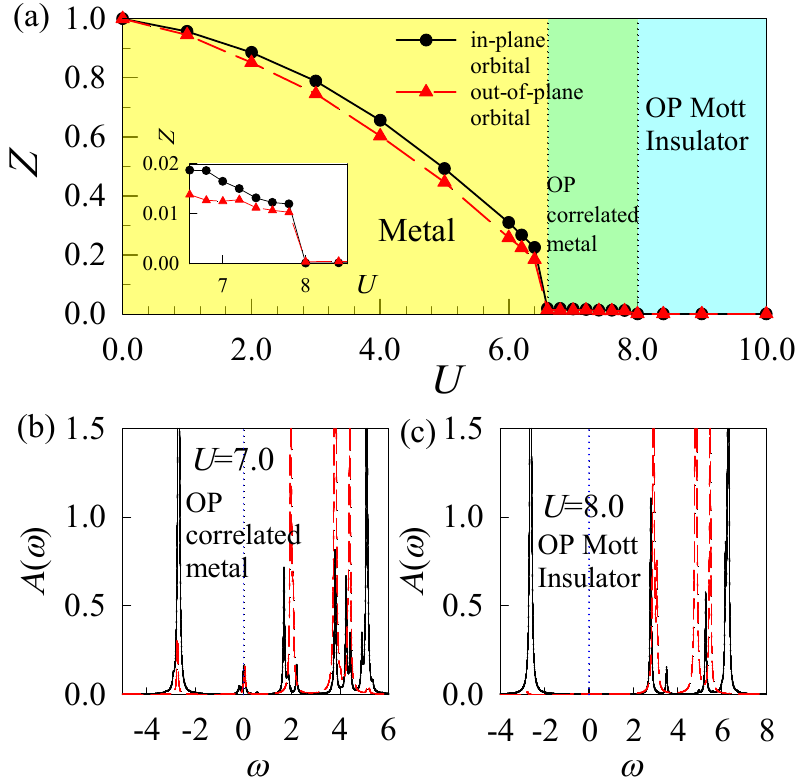}
\caption{Correlation-driven evolution from an OP correlated metal to an OP Mott insulator at quarter filling for $J_{\rm H}=0.05U$.
(a) Orbital-resolved quasiparticle weights $Z_l$ as functions of $U$.
The inset highlights that $Z_l$ remains finite after orbital polarization emerges, demonstrating that orbital reconstruction precedes Mott localization and produces an orbital-polarized (OP) correlated metallic phase.
The vanishing of $Z_l$ near $U=8.0$ signals the Mott transition.
(b) Orbital-resolved spectral functions at $U=7.0$, corresponding to the OP correlated metal.
(c) Orbital-resolved spectral functions at $U=8.0$, corresponding to the OP Mott insulator.
\label{Z_DOS_PT}}
\end{figure}

Once the interaction strength exceeds the orbital-reconstruction transition region, the system enters a strongly correlated OP state. To characterize the accompanying localization behavior, we calculate the quasiparticle weights and spectral functions.
As shown in Fig.~\ref{Z_DOS_PT}(a), $Z_l$ decreases continuously with increasing $U$.
After the orbital-reconstruction transition ($U\approx6.6$), the quasiparticle weights are strongly reduced but remain finite, as emphasized by the inset.
The finite quasiparticle coherence indicates that the OP state initially emerges as a correlated metallic phase rather than an insulating state\cite{Georges1996,Kotliar2006,Georges2013Ann}.
Only near $U=8.0$ do the quasiparticle weights vanish, indicating a Mott transition occurring after the establishment of the OP phase.
Importantly, the orbital-reconstruction transition occurs before the onset of Mott localization, and the subsequent metal-insulator transition takes place within an already established OP state. This demonstrates that orbital reconstruction is not merely a consequence of electron localization; rather, the Mott insulating phase emerges from a pre-existing OP electronic state and is accompanied by strong orbital reconstruction.

The spectral functions confirm this distinction.
At $U=7.0$ [Fig.~\ref{Z_DOS_PT}(b)], finite spectral weight remains at the
Fermi level, whereas at $U=8.0$ [Fig.~\ref{Z_DOS_PT}(c)] a clear gap opens and
the electrons become predominantly localized in the in-plane orbital.
Unlike a conventional Mott transition occurring in an orbitally balanced
system, the localization here develops after orbital reconstruction has
already taken place, leading to an OP Mott insulating phase.

Such correlation-enhanced localization is relevant to low-dimensional vanadium oxides, where reduced dimensionality simultaneously modifies the bandwidth and orbital occupations \cite{Daniel2019,Nekrasov2006}.
In CaVO$_3$ thin films, dimensional confinement reduces the effective bandwidth and enhances electronic correlations, driving the system from a correlated metallic state toward an insulating phase \cite{Daniel2019,Beck2020}.
In related SrVO$_3$ thin films, dimensional confinement has been shown to induce preferential occupation of the in-plane orbital within the $t_{2g}$ manifold\cite{Yoshimatsu2010}.
Within the present framework, this situation corresponds to the evolution from an OP correlated metal to an OP Mott insulator as the effective interaction strength $U$ increases.
Therefore, the metal-insulator transition obtained here does not represent a conventional Mott transition in an orbitally balanced system, but rather a localization transition emerging from a pre-existing anomalous OP state.



\subsection{Hund's-coupling-dependent evolution of orbital-polarized phases at quarter filling}

\begin{figure}
\includegraphics[scale=0.50]{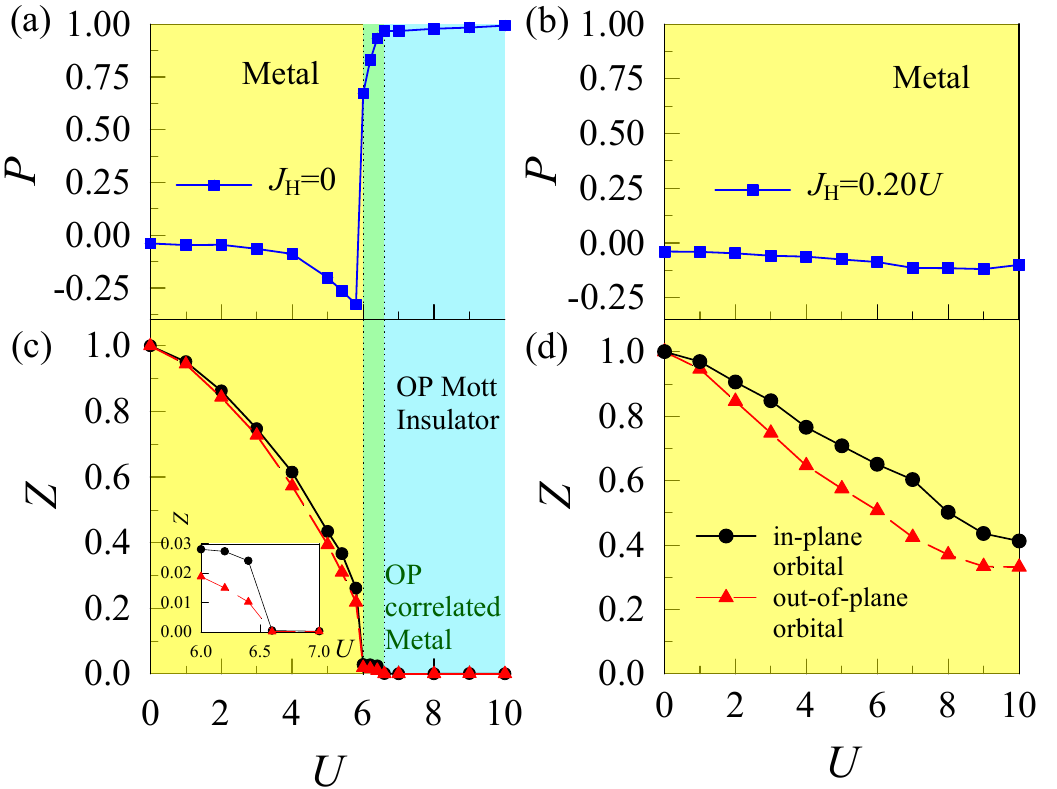}
\caption{Hund's coupling dependence of orbital reconstruction and quasiparticle coherence in the quarter-filled system.
(a) Orbital polarization order parameter $P$ as a function of $U$ for $J_{\rm H}=0$.
The orbital-reconstruction transition occurs at $U\approx6.0$, where $P$ rapidly increases from a weakly polarized metallic state to an OP state.
(b) Evolution of $P$ for strong Hund's coupling ($J_{\rm H}=0.20U$).
The orbital polarization is strongly suppressed and remains nearly absent over the whole interaction range.
(c) Quasiparticle weights $Z_l$ for $J_{\rm H}=0$.
The vanishing of $Z_l$ identifies the transition from an OP correlated metal phase to an OP Mott insulating phase.
The inset highlights the finite quasiparticle weight in the correlated metal regime.
(d) Quasiparticle weights for $J_{\rm H}=0.20U$.
The finite values of $Z_l$ indicate that the system remains metallic despite strong electronic correlations.
\label{Hund_QF}}
\end{figure}

To clarify the role of Hund's coupling in the orbital-reconstruction transition, we first examine the quarter-filled system without Hund's coupling. Figure~\ref{Hund_QF}(a) shows the evolution of the orbital polarization order parameter $P$ for $J_{\rm H}=0$.
Compared with the finite-Hund case discussed previously, the orbital-reconstruction transition occurs at a smaller interaction strength. As $U$ increases, $P$ remains close to zero in the metallic regime and then exhibits a sharp increase around $U\approx6.0$, indicating a transition into a strongly OP state. The corresponding quasiparticle weights shown in Fig.~\ref{Hund_QF}(c) demonstrate that the OP state initially appears as a correlated metal phase, where the quasiparticle weights are strongly reduced but remain finite.
Upon further increasing $U$, both quasiparticle weights vanish around $U\approx6.6$, accompanied by the opening of a Mott gap. Therefore, the system undergoes a successive transition from a correlated metallic phase to an OP correlated metal phase and finally to an OP Mott insulating phase.

We next investigate the effect of a strong Hund's coupling with $J_{\rm H}=0.20U$. As shown in Fig.~\ref{Hund_QF}(b), the orbital polarization is almost completely suppressed over the entire interaction range. Unlike the weak-Hund case, increasing $U$ no longer induces a significant transfer of electrons between the two orbitals.
The corresponding quasiparticle weights in Fig.~\ref{Hund_QF}(d) remain finite even at large interaction strengths, indicating that the system stays metallic. Although electronic correlations strongly reduce quasiparticle coherence, Hund's coupling prevents the development of the OP insulating state.
This behavior reveals that Hund's coupling competes directly with the interaction-driven orbital reconstruction mechanism.
While the interorbital Coulomb interaction $U'$ favors orbital redistribution
through enhanced interorbital charge competition, Hund's coupling tends to
maintain a more balanced occupation of different orbitals by favoring
high-spin configurations.
The finite quasiparticle coherence at large Hund's coupling reflects the well-known tendency of Hund's coupling to stabilize metallic states by suppressing orbital fluctuations and delaying localization \cite{Haule2009,Yin2011}. In the present system, however, the important role of Hund's coupling is not merely to preserve metallicity, but to compete with the interorbital charge redistribution responsible for orbital polarization.

The suppression of orbital polarization by Hund's coupling can be understood from its influence on interorbital charge fluctuations. The orbital-reconstruction transition discussed above relies on the increasing importance of the interorbital Coulomb interaction $U'$ after spectral reconstruction. Hund's coupling counteracts this tendency by favoring parallel-spin configurations and reducing the energetic advantage associated with transferring electrons into a single orbital\cite{Werner2007}.
Therefore, Hund's coupling does not simply shift the Mott transition boundary, but fundamentally modifies the stability of the OP phases. This competition between $U'$-driven orbital reconstruction and Hund-induced orbital balancing tendency determines the phase evolution of the quarter-filled system\cite{Medici2011B,Georges2013Ann}.

\begin{figure}
\includegraphics[scale=0.62]{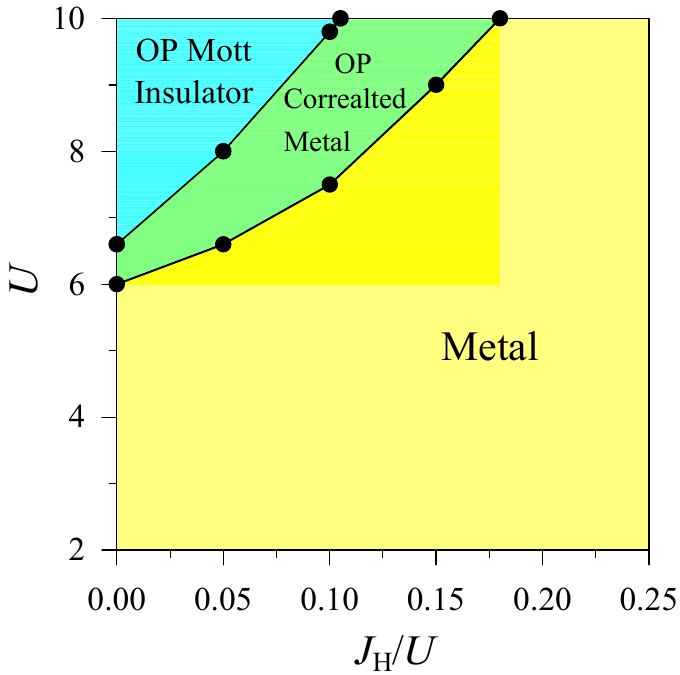}
\caption{Phase diagram of the quarter-filled system in the $U$--$J_{\rm H}/U$ plane.
The phase boundaries are determined from the evolution of orbital polarization parameter order and quasiparticle weights.
Three different phases are identified: metallic phase, OP correlated metal phase, and OP Mott insulating phase.
Increasing Hund's coupling systematically shifts the orbital-polarization and localization boundaries toward larger interaction strengths and eventually suppresses all OP phases.
\label{PG-QF}}
\end{figure}

The resulting phase diagram is summarized in Fig.~\ref{PG-QF}.
Both the orbital-polarization and Mott boundaries shift toward larger $U$
as $J_{\rm H}/U$ increases, and the OP phases eventually
disappear.
Thus, at quarter filling, Hund's coupling competes with the
$U'$-driven orbital-redistribution tendency and stabilizes the
metallic state.
The OP correlated metal and Mott insulator are consequently restricted to the strong-correlation and weak-Hund regime.
Such Hund's-coupling-dependent orbital redistribution is relevant to $d^1$ vanadium oxides such as SrVO$_3$ and CaVO$_3$ thin films, where dimensional confinement modifies the $t_{2g}$ orbital occupations and enhances orbital differentiation\cite{Yoshimatsu2010,Daniel2019,Beck2018}.




\subsection{Filling-dependent Hund's coupling physics at half filling}

\begin{figure}
\includegraphics[scale=0.60]{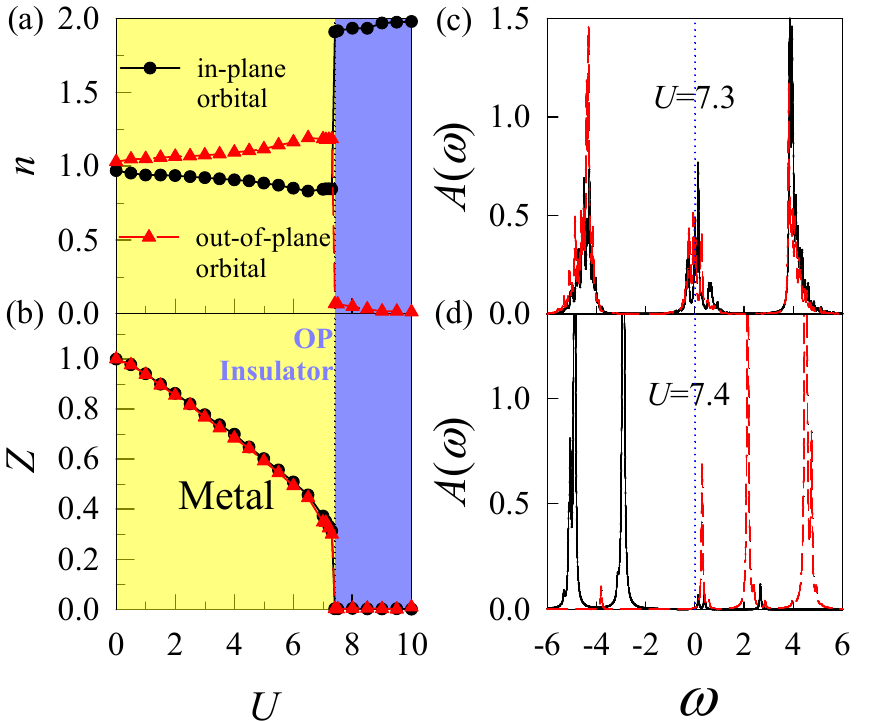}
\caption{Correlation-driven orbital reconstruction in the half-filled system without Hund's coupling ($J_{\rm H}=0$).
(a) Orbital-resolved occupations as a function of $U$.
The two orbitals exhibit nearly balanced occupations in the metallic regime, while a sudden orbital redistribution occurs at $U\approx7.4$, leading to a strongly OP insulating state.
(b) Quasiparticle weights $Z_l$ as a function of $U$.
The vanishing of $Z_l$ identifies the transition from a metallic phase to an OP insulating phase.
(c) and (d) Orbital-resolved spectral functions below and above the transition point, corresponding to $U=7.3$ and $U=7.4$, respectively.
The opening of the gap and disappearance of quasiparticle coherence confirm the emergence of an OP insulating phase.
\label{OP_HF}}
\end{figure}

The distinct role of Hund's coupling at different fillings raises an important question of how orbital reconstruction evolves away from quarter filling.
We therefore next investigate the half-filled system, where the competition between orbital polarization and Hund's coupling becomes qualitatively different from that in the quarter-filled case.
At half filling, Hund's coupling strongly favors high-spin configurations with electrons distributed among different orbitals, making the stability of orbital-polarized states particularly sensitive to the balance between interorbital charge transfer and Hund's coupling\cite{Georges2013Ann,Medici2011L}.

We first consider the case without Hund's coupling ($J_{\rm H}=0$). As shown in Fig.~\ref{OP_HF}(a), the two orbitals exhibit comparable occupations in the weakly correlated metallic regime. With increasing interaction strength, the occupation difference remains small until $U\approx7.4$, where a sudden orbital redistribution occurs. Electrons are transferred into the in-plane orbital, resulting in a strongly OP state despite its higher crystal-field energy, which is consistent with anomalous orbital polarization at quarter filling.
The corresponding quasiparticle weights shown in Fig.~\ref{OP_HF}(b) demonstrate that this orbital redistribution is accompanied by a metal-insulator transition. Both quasiparticle weights decrease continuously with increasing $U$ and vanish at the transition point. The spectral functions in Fig.~\ref{OP_HF}(c) and Fig.~\ref{OP_HF}(d) further confirm this behavior. Below the transition, the system exhibits a metallic state with finite spectral weight at the Fermi level, while above the transition a gap opens and the system enters an OP insulating phase.
Therefore, unlike the quarter-filled OP Mott phase, the insulating state at half filling does not originate from the localization of carriers within a single orbital channel.
Instead, it results from the simultaneous development of strong orbital polarization and interaction-induced insulating behavior.
We therefore denote this phase as an OP insulating phase.

\begin{figure}
\includegraphics[scale=0.60]{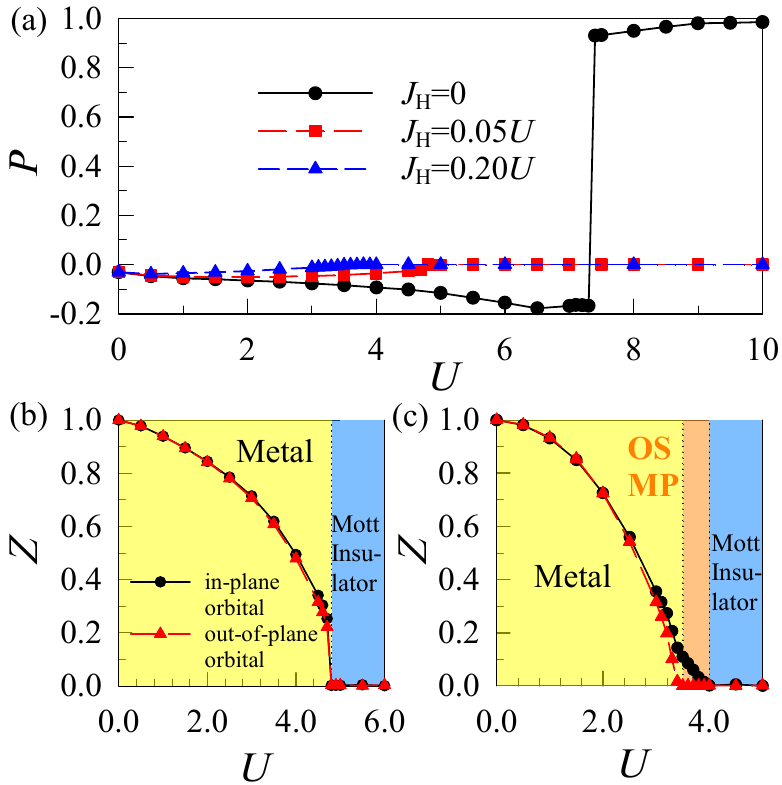}
\caption{Effect of Hund's coupling on orbital reconstruction and localization at half filling.
(a) Orbital polarization order parameter $P$ as a function of $U$ for different Hund's coupling strengths.
A sharp orbital-reconstruction transition appears only for $J_{\rm H}=0$, whereas finite Hund's coupling strongly suppresses the development of orbital polarization.
(b) Quasiparticle weights for $J_{\rm H}=0.05U$.
The system undergoes a conventional Mott transition from a metal to a Mott insulating state.
(c) Quasiparticle weights for $J_{\rm H}=0.20U$.
An orbital-selective Mott phase emerges between the metallic and Mott insulating phases, where one orbital becomes localized while the other remains itinerant.
\label{Hund_HF}}
\end{figure}

To clarify the effect of Hund's coupling at half filling, we examine the evolution of the orbital polarization and quasiparticle weights for finite Hund's coupling.
Figure~\ref{Hund_HF}(a) shows the orbital polarization order parameter for different values of $J_{\rm H}$. In contrast to the quarter-filled case, where Hund's coupling mainly shifts the orbital-polarization boundary, even a small finite Hund's coupling strongly suppresses the orbital-reconstruction transition at half filling.
This strong sensitivity originates from the special role of Hund's coupling at half filling. At half filling, each orbital tends to be singly occupied, and Hund's coupling strongly favors high-spin configurations with electrons distributed among different orbitals\cite{Werner2007,Medici2011L}. Therefore, the energy gain from maintaining orbital-balanced configurations competes directly with the $U'$-driven orbital reconstruction mechanism. As a result, finite Hund's coupling stabilizes the conventional Mott state and prevents electrons from collapsing into a single orbital\cite{Werner2007,Georges2013Ann}.
For $J_{\rm H}=0.05U$, the quasiparticle weights shown in Fig.~\ref{Hund_HF}(b) vanish simultaneously at the transition, indicating a conventional Mott transition. When Hund's coupling is further increased to $J_{\rm H}=0.20U$, the two orbitals exhibit different localization behaviors, as shown in Fig.~\ref{Hund_HF}(c). The orbital-dependent quasiparticle weights lead to an intermediate orbital-selective Mott phase, where one orbital becomes localized while the other remains metallic\cite{Medici2009,Georges2013Ann}.
The strong filling dependence of Hund's coupling is also relevant to multiorbital $d^2$ systems and related high-spin $d^3$ transition-metal oxides, where Hund's coupling plays a central role in stabilizing high-spin states and competing with orbital ordering tendencies\cite{De2007,Oles2009}.

\begin{figure}
\includegraphics[scale=0.60]{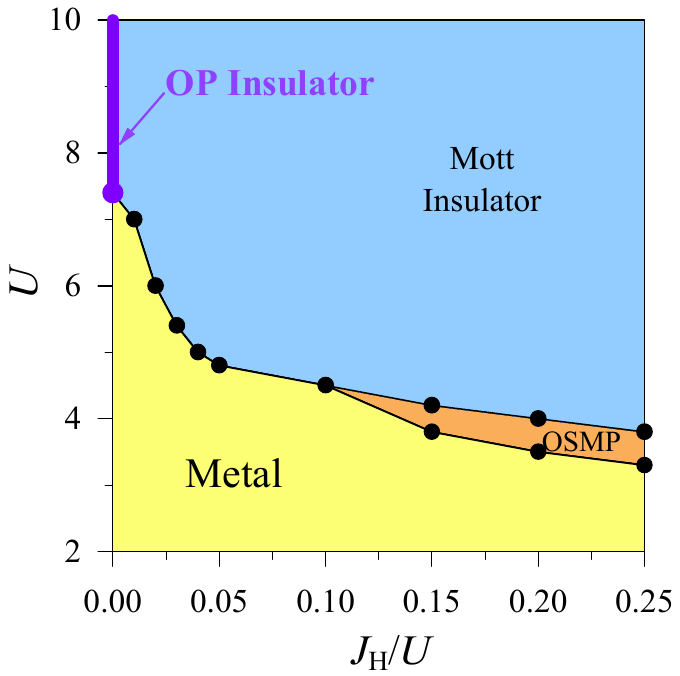}
\caption{Phase diagram of the half-filled system in the $U$--$J_{\rm H}/U$ plane.
The OP insulating phase exists only along the $J_{\rm H}=0$ axis and is immediately suppressed by finite Hund's coupling.
For finite $J_{\rm H}$, the system evolves from a metallic phase to a conventional Mott insulating phase, with an orbital-selective Mott phase emerging at sufficiently large Hund's coupling.
The phase diagram highlights the dominant role of Hund's coupling physics at half filling and its strong contrast with the quarter-filled case.
\label{PG-HF}}
\end{figure}


The complete phase diagram at half filling is summarized in Fig.~\ref{PG-HF}. The most striking feature is the extreme fragility of the OP insulating phase. Unlike the quarter-filled system, where orbital polarization survives within a finite range of Hund's coupling, the half-filled OP phase is restricted exclusively to $J_{\rm H}=0$.
Once Hund's coupling is introduced, the phase boundary of the OP state collapses immediately. The system instead undergoes a conventional Mott transition, and for sufficiently strong Hund's coupling an orbital-selective Mott phase emerges between the metallic and insulating states.
These results indicate that Hund's coupling plays fundamentally different roles depending on electron filling.
At quarter filling, Hund's coupling acts as a competing interaction that weakens orbital polarization. At half filling, the high-spin energy gain associated with Hund's coupling becomes maximized, strongly favoring interorbital spin alignment and suppressing charge redistribution between orbitals. Consequently, Hund's coupling fundamentally changes the nature of the correlated insulating state and prevents the formation of OP phases.

These filling-dependent effects of Hund's coupling highlight the need for a framework that can capture both correlation-driven orbital reconstruction and the role of dimensionality.
The mixed-dimensional framework developed here provides a minimal yet physically motivated platform for exploring orbital reconstruction in low-dimensional transition-metal oxides, including oxide heterointerfaces and thin films.
Our results demonstrate that such orbital reconstruction can lead to anomalous orbital polarization and exhibit a strongly filling-dependent response to Hund's coupling, providing a general mechanism beyond the conventional crystal-field picture.
The predicted asymmetric filling dependence of Hund's coupling offers a direct route for experimental verification.
For example, angle-resolved photoemission spectroscopy (ARPES) measurements on artificially confined oxide superlattices or ultrathin transition-metal oxide films could probe the evolution of orbital occupations, orbital-selective spectral weight transfer, and the emergence of OP metallic states prior to Mott localization.
These results suggest that controlling dimensionality and carrier filling provides a promising strategy for manipulating orbital reconstruction and correlated phases in oxide quantum materials.

\section{conclusion}

In summary, we have uncovered the microscopic mechanism of anomalous orbital
reconstruction and its interplay with electronic localization in a
mixed-dimensional two-orbital Hubbard model using DMFT. Motivated by orbital differentiation in low-dimensional transition
metal oxides and oxide interfaces, the model incorporates crystal-field
splitting, orbital-dependent dimensionality, and local Coulomb interactions
to capture the essential competition among different orbital channels.

Our results demonstrate that anomalous orbital polarization does not arise
from the bare crystal-field hierarchy alone. Instead, orbital-dependent
dimensionality and electronic correlations cooperate to reconstruct the
orbital occupations, with interorbital Coulomb interaction providing a key
microscopic channel through enhanced charge competition between orbitals.
The suppression of interorbital double occupancy and the enhancement of
negative interorbital charge correlation reveal how the correlated system
overcomes the initial crystal-field preference and transfers electrons into
the higher crystal-field-energy in-plane orbital.

At quarter filling, this orbital reconstruction first emerges within a
correlated metallic state and subsequently evolves into an OP
Mott insulating phase. This sequence demonstrates that orbital
reconstruction precedes localization, and that the resulting Mott state is
qualitatively different from a conventional orbitally balanced Mott phase.
The localized electrons retain a strong orbital polarization, indicating that the orbital reconstruction established in the metallic regime directly determines the nature of the subsequent Mott insulating phase.

Hund's coupling further acts as a filling-dependent regulator of orbital
reconstruction. At quarter filling, it weakens the correlation-induced
orbital redistribution by competing with interorbital charge fluctuations
and shifts the orbital-polarization and localization boundaries toward
stronger interactions. In sharp contrast, at half filling, Hund's coupling
completely suppresses the OP state by stabilizing high-spin,
orbital-balanced configurations.

Our results establish three key aspects of correlation-driven orbital reconstruction. First, orbital-dependent dimensionality and electronic correlations cooperate to reconstruct the orbital hierarchy, with interorbital Coulomb correlations providing the microscopic channel for orbital redistribution, clarifying the microscopic origin of anomalous orbital polarization beyond the crystal-field picture.
Second, orbital reconstruction emerges before Mott localization, leading to an OP correlated metallic phase. Third, Hund's coupling acts as a filling-dependent control parameter that determines whether orbital polarization survives or is replaced by conventional high-spin correlated phases.
These findings provide a general framework for understanding orbital reconstruction in low-dimensional multiorbital oxides.

\begin{acknowledgments}

This work was supported by the National Natural Science Foundation of China (Grants No. 12474062), Yunnan Fundamental Research Projects (Grant No. 202501AT070012) and Natural Science Foundation of Inner Mongolia Autonomous Region under Grant No. 21200-52531050.

\end{acknowledgments}

\appendix

\renewcommand{\thefigure}{\thesection\arabic{figure}}
\setcounter{figure}{0}

\section{Calculation of physical quantities}
\label{App-A}


In this appendix, we detail the calculation methods for the physical quantities presented in the main text, including the orbital-resolved spectral function, orbital occupation, quasiparticle weight, and the bath discretization scheme within the {\it Lanczos}-based DMFT framework.

The impurity problem is solved via exact diagonalization (ED), where the continuous hybridization function is approximated by a finite set of bath levels. For this purpose, we assign three bath sites per orbital. This bath discretization has been
shown to be sufficient for capturing the essential low-energy correlation
effects and phase transitions in multiorbital Hubbard models
\cite{Caffarel1994,Liebsch2006,Capone2007,Liebsch2012}. In particular,
previous {\it Lanczos}-DMFT studies of two-orbital Hubbard models demonstrated that increasing the bath size from $n_b=3$ to $n_b=4$ yields negligible changes in the low-energy self-energy, supporting the notion of an effectively doubled bath size near the orbital-selective Mott transition\cite{YKNiu2019}.

Employing the {\it Lanczos} solver, we can obtain the diagonal matrix element of the Green's function $G^{imp}_{l}$\cite{Dagotto1994,Caffarel1994,Capone2007,YKNiu2019},
which is expressed as
\begin{eqnarray}
G^{imp}_{l}\left( i \omega_n \right)=G_{l}^{(+)}\left( i \omega_n\right)+G_{l}^{(-)}\left( i \omega_n\right),
\end{eqnarray}
where
\begin{eqnarray}
G_{l}^{(+)}\left( i \omega_n \right)=\frac{\left\langle\phi_{0}\left|d_l d_l^{\dagger}\right| \phi_{0}\right\rangle}{  i \omega_n -a_{0}^{(+)}-\frac{b_{1}^{(+) 2}}{ i \omega_n -a_{1}^{(+)}-\frac{b_{2}^{(+) 2}}{  i \omega_n -a_{2}^{(+)}-\ldots}}},
\end{eqnarray}
and
\begin{eqnarray}
G_{l}^{(-)}\left(  i \omega_n \right)=\frac{\left\langle\phi_{0}\left|d_l^{\dagger} d_l\right| \phi_{0}\right\rangle}{  i \omega_n +a_{0}^{(-)}-\frac{b_{1}^{(-) 2}}{ i \omega_n +a_{1}^{(-)}-\frac{b_{2}^{(-) 2}}{  i \omega_n +a_{2}^{(-)}-\ldots}}}.
\end{eqnarray}

We build the DMFT self-consistent loop with
${G}^{lat}_{l}(i \omega _n)=G^{imp}_{l}(i\omega _n)$ to determine the parameters
$\varepsilon_{ml}$ and $V_{lm}$.
Analytic continuation is also performed to obtain the real frequency
Green's function ${ G(\omega)}$\cite{Georges1996,Kotliar2006}.

The orbital occupation and correlation functions are directly evaluated from the many-body ground state($\phi_0$) obtained from the {\it Lanczos} impurity solver,
\begin{equation}
n_l=
\langle\phi_0|\hat n_l|\phi_0\rangle ,
\end{equation}
and
\begin{equation}
D_{12}
=
\langle\phi_0|\hat n_1\hat n_2|\phi_0\rangle .
\end{equation}
These quantities are calculated from the ground-state wave function to avoid numerical uncertainties associated with the finite broadening used in the real-frequency spectral functions.

The occupation obtained from the spectral function integration,
\begin{equation}
n_l=
2\int A_l(\omega)f(\omega)d\omega ,
\end{equation}
is consistent with the above definition within numerical accuracy.

The quasiparticle weight is used to characterize the coherence of the correlated electronic states.
For each orbital, it is defined as
\begin{equation}
Z_l
=
\left(
1-
\frac{\partial {\rm Re}\Sigma_l(\omega)}
{\partial\omega}
\bigg|_{\omega=0}
\right)^{-1}.
\end{equation}
In practical calculations, the self-energy is evaluated on the imaginary frequency axis.
Therefore, the quasiparticle weight is obtained using the low-frequency approximation,
\begin{equation}
Z_l\simeq\left(1-\frac{{\rm Im}\Sigma_l(i\omega_0)}{\omega_0}\right)^{-1},
\end{equation}
where $\omega_0$ is the lowest Matsubara frequency.
A finite value of $Z_l$ indicates the existence of coherent quasiparticle excitations, whereas $Z_l\rightarrow0$ signals the disappearance of quasiparticle coherence and the formation of a Mott insulating state.

\section{Origin of the nonmonotonic orbital occupation in the metallic regime}
\label{App-B}

\begin{figure}
\includegraphics[scale=0.56]{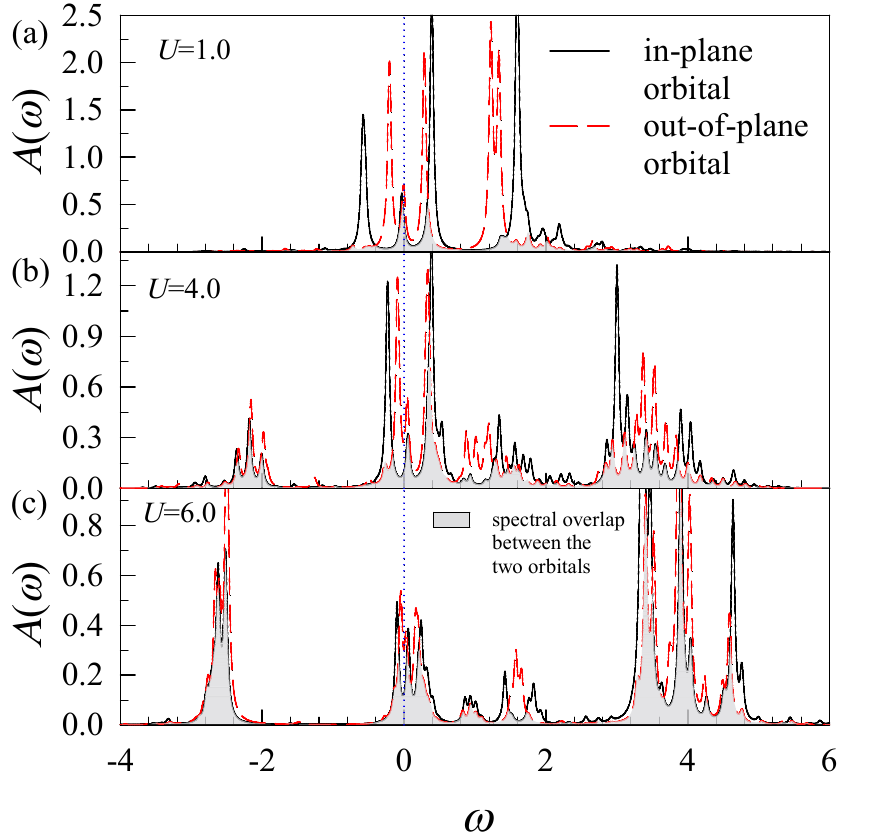}
\caption{Orbital-resolved spectral functions in the metallic regime for different interaction strengths.
(a) $U=1$, (b) $U=4$, and (c) $U=6$.
Increasing interaction strength enhances the spectral overlap between the two orbitals and modifies their low-energy spectral weights, explaining the nonmonotonic orbital occupation evolution before the orbital-reconstruction transition.
\label{DOS-metal}}
\end{figure}

In the metallic regime, the orbital occupations exhibit a nonmonotonic evolution with increasing interaction strength.
To clarify this behavior, we analyze the orbital-resolved spectral functions.

The orbital-resolved spectral density is obtained from the lattice Green's function as
\begin{equation}
A_l(\omega)
=
-\frac{1}{\pi}
\mathrm{Im}
G^{\mathrm{lat}}_l(\omega+i\eta),
\end{equation}
where $l$ denotes the orbital index and $\eta$ is a small energy broadening factor.
In the present calculations, we take $\eta=0.02$.

Since we consider a paramagnetic system, the spectral functions for the two spin components are identical,$A_{l\uparrow}(\omega)=A_{l\downarrow}(\omega)$,
and therefore we use $A_l(\omega)$ to represent the spin-independent orbital-resolved spectral function.

Figures~\ref{DOS-metal}(a)-(c) show the spectral functions for different interaction strengths.
Increasing $U$ leads to a pronounced spectral-weight redistribution.
For weak interaction ($U=1$), both orbitals exhibit well-defined quasiparticle peaks near the Fermi level, and the orbital occupations are mainly determined by the crystal-field splitting.
Upon increasing $U$, correlation effects gradually reconstruct the electronic spectra.
Because the out-of-plane orbital has a lower effective dimensionality and a reduced hopping phase space, its lower Hubbard band develops more rapidly than that of the in-plane orbital, resulting in an asymmetric redistribution of spectral weight.
Consequently, additional spectral weight is transferred toward low-energy states of the out-of-plane orbital, explaining the increase of its occupation in the correlated metallic regime.

It is also worth noting that the spectral overlap between the two orbitals is enhanced with increasing interaction strength.
Although this overlap is not the primary origin of the nonmonotonic occupation behavior in the metallic regime, it increases the relevance of interorbital charge fluctuations and provides the basis for the subsequent orbital-polarization instability discussed in the main text.

\section{Effect of orbital dimensionality on orbital reconstruction}
\label{App-C}

\begin{figure}
\includegraphics[scale=0.62]{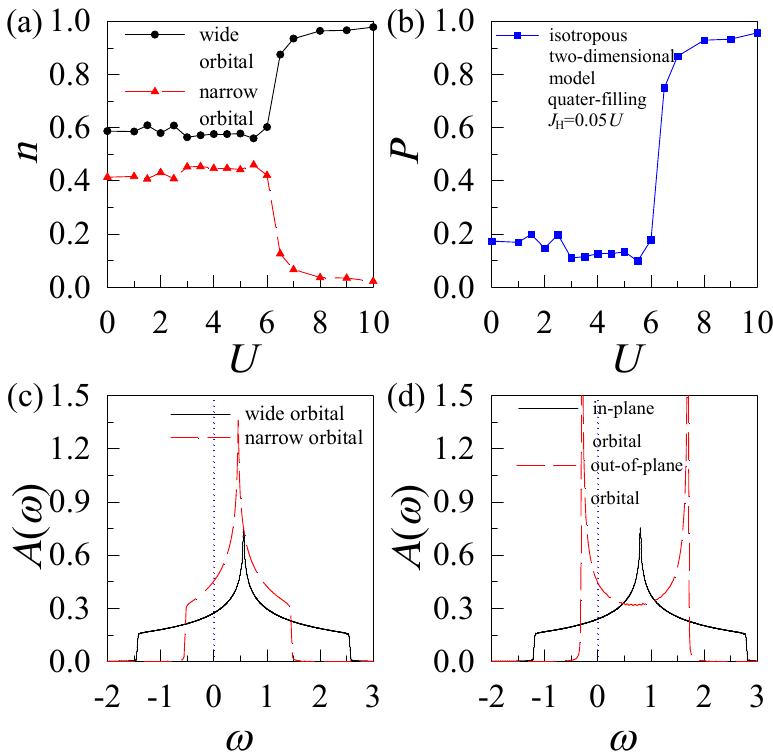}
\caption{Effect of orbital dimensionality on orbital reconstruction.
(a) Orbital occupations and (b) orbital polarization for the isotropic two-dimensional narrow-band model.
(c) Noninteracting density of states for the isotropic two-dimensional model and (d) the one-dimensional orbital model used in the main text.
The absence of a low-energy spectral peak in the narrow two-dimensional orbital prevents occupation inversion, although the interaction-driven orbital reconstruction mechanism remains robust.
\label{orbital_dimensionality}}
\end{figure}

In the main text, orbital 2 is chosen as a one-dimensional orbital to mimic the reduced dimensionality of out-of-plane orbitals in low-dimensional transition-metal oxides.
To examine whether the anomalous orbital polarization depends sensitively on this specific dimensionality, we further consider an isotropic two-dimensional narrow-band orbital model, where
$t_{1x}=t_{1y}=0.5$
and
$t_{2x}=t_{2y}=0.25$.

As shown in Fig.~\ref{orbital_dimensionality}(a), the isotropic two-dimensional
model still exhibits an interaction-driven orbital reconstruction transition.
However, in contrast to the mixed-dimensional case discussed in the main text,
the orbital occupations do not invert before the transition, and the orbital
polarization remains positive. This indicates that the correlation-induced
orbital reconstruction mechanism is robust, while the anomalous occupation
inversion is sensitive to the orbital-dependent dimensionality.

The evolution of the orbital polarization in
Fig.~\ref{orbital_dimensionality}(b) further clarifies this distinction.
With increasing interaction strength, $P$ remains positive in the metallic
regime and increases abruptly near the transition, indicating that the system
develops a preferential occupation of the wide in-plane orbital. Therefore,
the absence of occupation inversion in the isotropic two-dimensional model
does not imply the absence of correlation-driven orbital reconstruction;
instead, it reflects a different balance between kinetic-energy and
interaction effects induced by orbital dimensionality.

To understand the origin of this difference, we compare the noninteracting
density of states of the two models.
For the isotropic two-dimensional model [Fig.~\ref{orbital_dimensionality}(c)],
the narrow orbital does not possess the pronounced low-energy quasiparticle
peak observed for the one-dimensional orbital
[Fig.~\ref{orbital_dimensionality} (d)]. Therefore, at quarter filling, the
larger bandwidth of the wide orbital dominates the orbital occupation, and
the chemical potential is insufficient to overcome this kinetic-energy
preference.

In contrast, the one-dimensional orbital exhibits enhanced low-energy
spectral weight, which leads to a larger occupation despite its narrower
bandwidth. Thus, the anomalous orbital occupation inversion is related to
the dimensionality-dependent spectral structure, whereas the
interaction-driven orbital-reconstruction mechanism itself remains robust.
Our findings also suggest bandwidth control as an alternative pathway for orbital polarization engineering in correlated oxides, in agreement with the inverted orbital polarization reported in CaFeO$_3$ thin films\cite{Rogge2018}.

\bibliography{OPT-HF_QF}

\end{document}